# Asteroid Thermal Inertia Estimates from Remote Infrared Observations: The Effects of Surface Roughness and Rotation Rate

Alan W. Harris[1] and Line Drube[1,2]

[1]German Aerospace Center (DLR) Institute of Planetary Research, Rutherfordstrasse 2, D-12489 Berlin, Germany; alan.harris@dlr.de

[2]Present address: Technical University of Denmark, Building 356, 2800 Kgs. Lyngby, Denmark

# Abstract

The thermal inertia of an asteroid's surface can provide insight into regolith properties, such as the presence of a layer of fine dust, the density and thermal conductivity of a rocky surface, and, together with other observational data, mineralogy. Knowledge of the surface characteristics of asteroids is important for planetary defense initiatives and the extraction of resources ("asteroid mining"). A simple means of estimating asteroid thermal inertia has been proposed by Harris & Drube, which is suitable for application to large sets of thermal-infrared observational data, such as those obtained by infrared space telescopes. We compare results from the Harris-Drube estimator with recently published values of asteroid thermal inertia from detailed thermophysical modeling, and provide an explanation in terms of reduced surface roughness for some discrepant results. Smooth surfaces covered in fine dust may provide an explanation for the unexpectedly low values of thermal inertia derived from thermophysical modeling for some slowly rotating main-belt asteroids (MBAs). In the case of near-Earth objects (NEOs) we show that results from the estimator are in good agreement with those from thermophysical modeling, with just a few exceptions. We discuss the special cases of the NEOs (101955) Bennu, (162173) Ryugu, and (29075) 1950 DA in the context of results from our estimator. Given the data requirements and complexity of thermophysical modeling, data-analysis tools based on relatively simple concepts can play an important role in allowing "quick-look" assessment of thermal-infrared data of asteroids, especially NEOs.

# 1. Introduction

Asteroids have become a focus of attention during recent decades due to the realization that they are remnants of objects that played a key role in planet formation, the fact that they may one day become sources of raw materials for future spacefaring generations, and the danger to our civilization posed by near-Earth objects (NEOs) that can cross the Earth's orbit. Much current research effort is dedicated to investigating the physical properties of asteroids via Earth-based telescopes, and space missions to flyby, rendezvous with, and return samples from, selected objects. We now have a wealth of information on parameters such as size, albedo, shape, spin state, and mineralogy for hundreds or thousands of asteroids but relatively little reliable information on density, porosity, subsurface and deep interior structure.

A survey of the sizes and albedos of more than 100,000 asteroids has been carried out by the NASA Wide-Field Infrared Survey Explorer space telescope (*WISE*, Wright et al. 2010), which was launched to Earth orbit in 2009 December carrying a 40 cm diameter telescope and infrared detectors. The *WISE/NEOWISE* program (Mainzer et al. 2011; Masiero et al. 2011) analyzed images collected during the cryogenic phase of the mission in up to four infrared bands, centered on 3.4, 4.6, 12, and 22 μm, to derive asteroid diameters, albedos, and best-fit values of a modeling parameter, $\eta$, often called the beaming parameter, related to the temperature distribution on the object's surface (Harris 1998).

Here we describe the results of a study of *WISE* and other thermal-infrared data in which we have derived estimates of asteroid thermal inertia from best-fit $\eta$ values. The thermal inertia of an asteroid's surface increases with increasing density and thermal conductivity; it provides a guide to its porosity and cohesion. Low values of thermal inertia are consistent with a dusty and/or porous regolith, while high values are indicative of low-porosity rocky material with relatively high thermal conductivity. Most current models of the thermal properties of atmosphereless bodies treat thermal inertia as a property of the surface and do not consider possible variation with depth. However, it appears that some results can be best explained in terms of an increase of thermal inertia with depth, including the steeper than expected dependence of thermal inertia on heliocentric distance for some objects (Rozitis et al. 2018), and the relatively low thermal-inertia values derived from observations of Jupiter satellites and Trojans made during eclipses (Mueller et al. 2010). The eclipse thermal-inertia values, based on measurements of surface temperature changes during shadowing, are factors of 4 or 5 lower than the diurnal values. During eclipse events, which are much shorter than the spin period, the thermal wave presumably does not penetrate deep enough to reach subsurface layers of higher thermal inertia. Harris & Drube (2016), on the basis of their thermal-inertia estimator, provided evidence for a dependence of remotely sensed thermal inertia on rotation period, which they interpreted in terms of deeper penetration of solar energy into subsolar surface elements of longer period objects (larger thermal skin depths – see Section 5), leading to enhanced absorption due to the deeper subsurface material having higher density and/or thermal conductivity.

However, recent results from Marciniak et al. (2019), on the basis of thermophysical modeling, find that some objects with long periods have very low thermal-inertia values, in contrast to expectations from the work of Harris & Drube (2016). Furthermore, results from thermophysical modeling of remote thermal-infrared observations of some main-belt asteroids (MBAs) are indicative of very low values of thermal inertia compared to those estimated from best-fit $\eta$ values (Hanuš et al. 2018; MacLennan & Emery 2019).

In contrast, the thermal-inertia values of rocks on the spacecraft targets (101955) Bennu, and (162173) Ryugu from in situ measurements are much higher than values associated with dusty, fine-grained material, but lower than expected for rocky material. While the in situ values are in broad agreement with those from ground-based observations, their interpretation in the absence of fine-grained insulating material requires an unexpected type of highly porous rock (Lauretta et al. 2019; Okada et al. 2020).

On the basis of published thermophysically modeled thermal-inertia values and results from our estimator, we explore the implications of thermal-inertia results for the surface properties of asteroids and show that there is a correlation between the shortfall in measured thermal inertia compared to estimates based on best-fit $\eta$ values and the smoothness of the surface, which provides an explanation for the discrepancy in terms of the assumptions regarding surface roughness made in the derivation of the thermal-inertia estimator of Harris & Drube (2016).

## 2. Calculating Asteroid Thermal Inertia

Calculating an accurate value of thermal inertia for an asteroid requires knowledge of the object's shape and spin state, and thermal-infrared observations made over a broad range of phase and aspect angles. These data serve as the input for a thermophysical model, which incorporates a shape model of the asteroid, and physical descriptions of its surface roughness, scattering and re-absorption of insolation, and thermal transport into the surface. The resulting thermal emission from the surface is compared to that deduced from the thermal-infrared observations; the thermophysical model parameters are adjusted in an iterative process that searches for the best fit of the model to the observational data. Thermophysical modeling is a complex but powerful means of providing reliable sizes, albedos, and thermal-inertia values of atmosphereless bodies. For historical context and further details, see, for example, Spencer (1990) and the reviews of Harris & Lagerros (2002) and Delbo' et al. (2015).

A simpler but less accurate approach to the derivation of thermal inertia has been proposed by Harris & Drube (2016), which is suitable for application to large sets of thermal-infrared observational data, such as those obtained by the *WISE/NEOWISE* survey, and objects for which little physical information is available. The Harris-Drube thermal-inertia estimator is based on an approximately linear relationship between best-fit $\eta$ values (see above) and the thermal parameter as defined by Spencer et al. (1989), which is a measure of the ability of the asteroid surface temperature to keep up with diurnal insolation changes and is proportional to thermal inertia. A similar method used by Harris et al. (1998) revealed that the NEOs (2100) Ra-Shalom and (3200) Phaethon have relatively high values of thermal inertia. Furthermore, Harris & Drube (2014) showed that $\eta$ appears to be a useful indicator of asteroids containing large amounts of metal (see also Moeyens et al. 2020).

For the reader's convenience we repeat here the expression from Harris & Drube (2016) for estimating thermal inertia, $\Gamma$, given a measurement of $\eta$:

$$\Gamma = (\eta_{norm} - \eta_{norm,0})((1-A)^{3/4}\varepsilon^{1/4}\sigma^{1/4}S_1^{3/4})/(b \sin\theta\, \omega^{1/2}R^{3/2}), \qquad (1)$$

where $\eta_{norm,0} = 0.74$, $b = 0.38$, $A$ is the bolometric Bond albedo, $\varepsilon$ is the emissivity, $S_1$ is the solar constant, $\theta$ is the angle between the object's spin axis and the solar direction, $\omega$ is the spin rate, $R$ is the heliocentric distance in astronomical units, and $\eta_{norm} = \eta + 0.00963\,(50°-\alpha°)$ (Mainzer et al. 2011) where $\alpha$ is the solar phase angle. There is little evidence that $\eta$ depends on $\alpha$ for $\alpha < 20°$ (Mainzer et al. 2011), therefore for the purposes of normalization we take $\alpha = 20°$ for phase angles below 20°. Since the Sun's spectral energy distribution peaks in the visible region and the dependence of asteroid albedos on wavelength is normally small, it is normal to assume that $A = A_V = q\, p_V$, where $A_V$ is the visual Bond albedo, and $q$ is the phase integral. This allows the physically significant parameter $A$ to be linked directly to the observationally derived visible geometric albedo, $p_V$. In the standard $H$, $G$ magnitude system (Bowell et al. 1989), in which $H$ is the absolute magnitude and $G$ the slope parameter, $q = 0.290+0.684G$.

In contrast to thermophysical modeling the estimator requires only knowledge of $\eta$ and the object's spin state. The thermal parameter increases with spin rate, and the inclination of the object's spin axis to the solar direction, $\theta$, has a major influence on the surface distribution of insolation. Harris & Drube (2016) give the accuracy of their estimator as a factor of two for (TP $\sin\theta$) in the range 0.75 – 3.5, where TP is the thermal parameter. We note that $\sin\theta$ is very often close to 1 (see Tables 1, 1A, and Harris & Drube, 2016, Table 1). In Figure 1 we plot previously published thermophysically modeled values of thermal inertia listed in Table A.2 of Hanuš et al. (2018) against values derived using the estimator. Objects with TP$_{TPM}$ $\sin\theta$ (derived from thermophysical modeling) in the range 0.75 – 3.5 and associated parameter values, which are additional to those listed in Harris & Drube (2016), are given in Table 1. Objects for which the thermal-parameter values are outside the limits of applicability of the estimator, i.e. those with (TP$_{TPM}$ $\sin\theta$) < 0.75 or > 3.5, are plotted as open red squares in Figure 1 and listed in Table A1. Many of the red-square points in Figure 1 lie within or close to a factor of 2 of the thermophysically derived values, despite their representing objects with thermal-parameter values outside the formal limits of applicability of the estimator. In total 78% of all plotted points lie within the region bounded by the dashed lines, in which agreement is good to within a factor of 2. If the count is restricted to only those points satisfying the criterion for applicability of the estimator (black points), the dashed lines enclose 92% of the points.

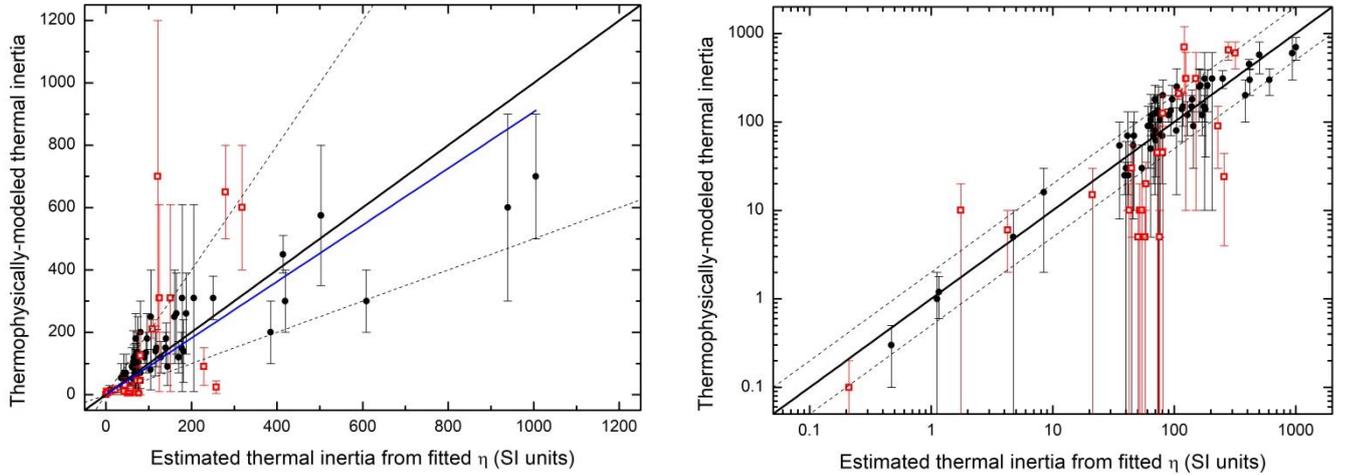

**Figure 1a (left)**. Thermal-inertia values derived from thermophysical modeling listed in the literature compilations of Delbo' et al. (2015) and Hanuš et al. (2018, Table A.2) compared to values derived from the estimator of Harris & Drube (2016). For comparison purposes values of $TI_{est}$, obtained using the Harris & Drube (2016) estimator, have been normalized to the heliocentric distances, $R$, given in Table A.2 of Hanuš et al. (2018) using the $R^{-3/4}$ scaling relation discussed by Delbo' et al. (2015). Objects for which the thermal-parameter values are outside the formal limits of applicability of the estimator, i.e. those with $(TP_{TPM} \sin\theta) < 0.75$ or $> 3.5$, are plotted as open red squares and listed in Table A1. The dashed lines represent the range in which the thermophysically derived thermal inertia is within a factor of 2 of the estimated value. Values of $\eta$ for an object derived from independent sets of data are treated as separate values, thus some objects are represented by more than one data point. The blue line is a weighted linear fit to the points with $0.75 < (TP_{TPM} \sin\theta) < 3.5$. The correlation coefficient is 0.93 and the slope is 0.91, i.e. the fit is close to the line of equality. SI units of thermal inertia are J m$^{-2}$s$^{-0.5}$K$^{-1}$. In the case of (162173) Ryugu a "ground-truth" value of thermal inertia derived by Okada et al. (2020) from Hayabusa 2 results has been taken ($\Gamma = 300 \pm 100$ at 1.1 au).

**Figure 1b (right)**. The data of Figure 1a plotted on a log scale to decompress the lower end of the thermal-inertia range. This plot is an updated version of that in Figure 4 of Harris & Drube (2016).

**Table 1**

Data in Figures 1, 2, and 7 Additional to Those Plotted in Harris & Drube (2016, Figures 4, 7) for which (TP$_{TPM}$ sin$\theta$) Is Within the Limits of Applicability of the Estimator (0.75-3.5)

| Name | G | R (au) | Period (h) | $p_V$ | $\alpha°$ | sin $\theta$ | $\eta$ | $\eta_{norm}$ | TP$_{TPM}$ sin$\theta$ | $\Gamma_{est}$ | $\Gamma_{TPM}$ | Data Source |
|---|---|---|---|---|---|---|---|---|---|---|---|---|
| 771 Libera | 0.15 | 3.1 | 5.892 | 0.12 | <20 | 0.97 | 1.05±0.01 | 1.34±0.01 | 2.34 | 61 | 90±60 | 1 |
| 771 Libera | 0.15 | 2.8 | 5.892 | 0.12 | 20.9 | 1.0 | 1.05±0.04 | 1.33±0.05 | 2.20 | 64 | 90±60 | 1 |
| 1036 Ganymed | 0.3 | 3.5 | 10.30 | 0.23 | <20 | 0.95 | 0.96±0.04 | 1.25±0.05 | 1.53 | 46 | 54±46 | 1 |
| 1036 Ganymed | 0.3 | 3.9 | 10.30 | 0.23 | <20 | 0.91 | 0.84±0.03 | 1.13±0.04 | 1.59 | 35 | 54±46 | 1 |
| 1472 Muonio | 0.15 | 2.7 | 8.706 | 0.24 | 22.1 | 0.96 | 1.02±0.03 | 1.28±0.04 | 0.84 | 74 | 44±45 | 1 |
| 1627 Ivar | 0.6 | 2.1 | 4.795 | 0.13 | 28.3 | 0.73 | 1.13±0.03 | 1.34±0.04 | 1.88 | 116 | 140±80 | 1 |
| 1685 Toro | 0.15 | 1.9 | 10.20 | 0.25 | 32.6 | 0.99 | 1.24±0.05 | 1.41±0.06 | 2.69 | 168 | 260±130 | 1 |
| 1685 Toro | 0.15 | 1.9 | 10.20 | 0.25 | 32.6 | 0.99 | 1.34±0.08 | 1.51±0.09 | 2.79 | 187 | 260±130 | 1 |
| 1980 Tezcatlipoca | 0.15 | 1.1 | 7.246 | 0.15 | 62.9 | 1.0 | 1.64±0.33 | 1.52±0.30 | 3.06 | 205 | 310±300 | 2 |
| 1980 Tezcatlipoca | 0.15 | 1.1 | 7.246 | 0.15 | 63.4 | 1.0 | 1.54±0.31 | 1.41±0.28 | 3.05 | 178 | 310±300 | 2 |

**Notes:** The listed thermophysically modeled thermal-inertia values, $\Gamma_{TPM}$, are from Hanuš et al. (2018, Table A.2). For comparison purposes the listed values of $\Gamma_{est}$, obtained using the Harris & Drube (2016) estimator, have been normalized to the heliocentric distances, $R$, given in Table A.2 of Hanuš et al. (2018) using the $R^{-3/4}$ scaling relation discussed by Delbo' et al. (2015). Units of thermal inertia are J m$^{-2}$s$^{-0.5}$K$^{-1}$. Other symbols are $G$: slope parameter, $p_V$: geometric visual albedo, $\alpha$: solar phase angle, $\eta$: beaming parameter, and $\theta$: angle between the object's spin axis and the solar direction. Data sources for $G$, $p_V$, $\eta$ are:

1. WISE catalog (Mainzer et al. 2019);
2. Harris & Davies (1999) (WISE data are available but result in (TP$_{TPM}$ sin$\theta$) > 3.5, i.e. outside the range of applicability of the estimator - see Table A1).

Observing geometry was taken from the NASA Jet Propulsion Laboratory Solar System Dynamics Horizons facility and asteroid spin vectors from the Asteroid Lightcurve Database (Warner et al. 2018).

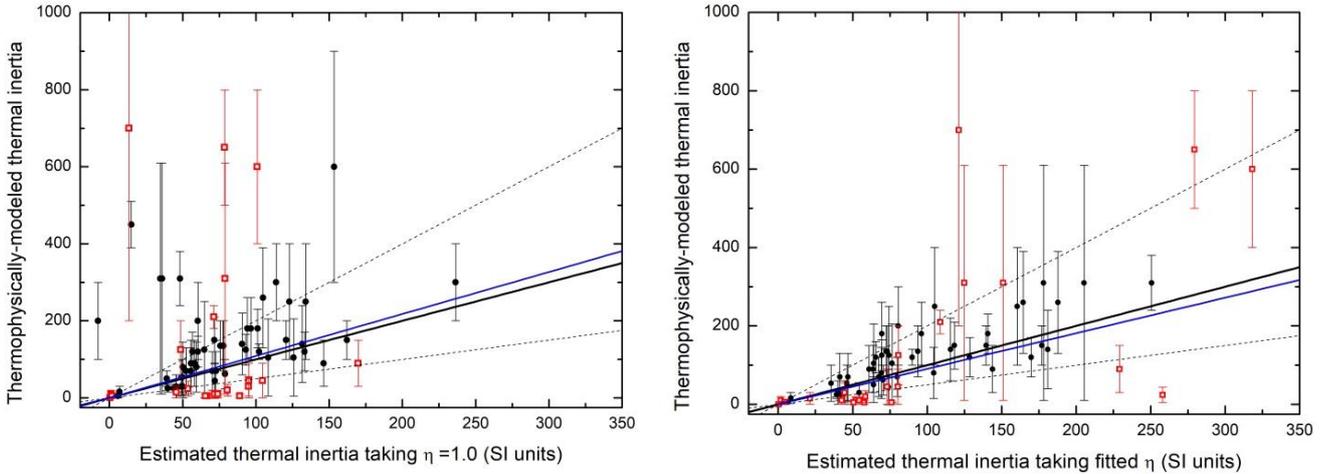

**Figure 2a (left).** Same as Figure 1, with $\eta$ treated as a constant equal to unity in the thermal-inertia estimator.
**Figure 2b (right).** Same as Figure 2a using the estimator as intended, i.e. taking fitted $\eta$. Comparison with Figure 2a demonstrates that fitted $\eta$ significantly reduces the scatter of the points. The correlation coefficient of the linear fit to the full data set plotted in Figure 1 (blue line) improves from 0.78 to 0.93 if fitted $\eta$ is taken instead of $\eta = 1$.

The $\eta$ fitting parameter is related to the temperature distribution on an object's surface and is important for deriving reliable sizes of atmosphereless bodies from thermal-infrared observations. The theory behind our thermal-inertia estimator has been discussed by Harris & Drube (2016) and is outlined above, but do $\eta$ values fitted to observational data really contain significant information on thermal inertia in practice? Given that for many objects, especially MBAs, $\eta$ is near to 1, would Equation (1) work just as well with $\eta = 1$ treated as a constant? In Figure 2a we plot the data of Figure 1, with the scale expanded in the lower thermal-inertia range for clarity, with $\eta = 1$ substituted for fitted $\eta$ in the estimator. In Figure 2b we plot estimated thermal inertia taking fitted $\eta$ for comparison. The fact that fitted $\eta$ contains significant information on thermal inertia is convincingly demonstrated by comparison of the frames in Figure 2.

### 3. Recently Published Asteroid Thermal-Inertia Data

Since the publication of Harris & Drube (2016) further sets of asteroid thermal inertia from thermophysical modeling have appeared in the literature. Hanuš et al. (2018, Table A.3) present results from fits using their "varied-shape thermophysical model" for over 100 asteroids. MacLennan & Emery (2019) present results derived from thermophysical modeling using ellipsoid shape models for 21 asteroids. Both studies are based on data from the *WISE* mission. We have derived thermal-inertia estimates for the objects in the above studies using the Harris-Drube estimator. The estimates are compared with the thermophysically derived thermal-inertia values in Figure 3, in which the filled symbols indicate that ($TP_{TPM} \sin\theta$) lies within the range 0.75 - 3.5 and open symbols indicate that ($TP_{TPM} \sin\theta$) lies outside this range, i.e. outside the range of applicability of the estimator, where $TP_{TPM}$ denotes the thermal parameter based on the thermophysically derived thermal inertia. The filled symbols lie within the expected factor of 2 uncertainty of the estimator (cf. Figure 2b), while most of the open symbols, especially those from the work of Hanuš et al. (2018), lie well below the line of equality, demonstrating that the estimator apparently produces gross overestimates in many of these cases.

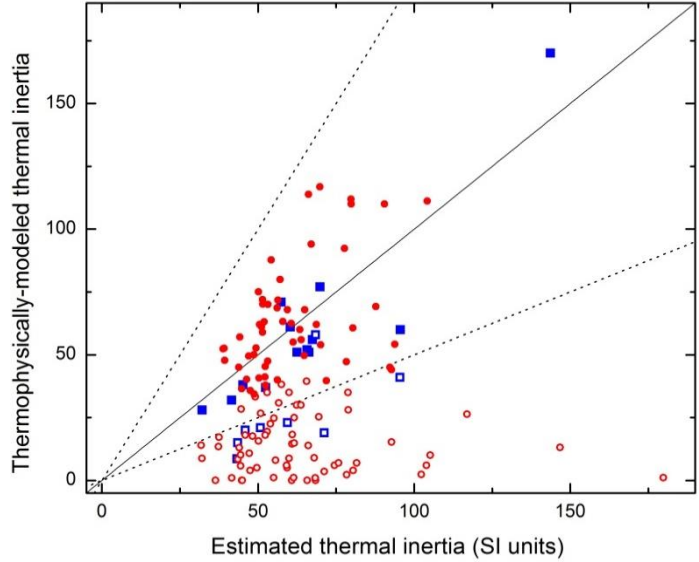

**Figure 3**. Thermal inertia from thermophysical modeling compared to values derived from the estimator of Harris & Drube (2016). Thermophysically modeled data from Hanuš et al. (2018, Table A.3) are plotted as red disks, those from MacLennan & Emery (2019) as blue squares. Note that 11 objects are common to both datasets. Objects for which the thermophysical values are outside the limits of applicability of the estimator, i.e. those with $(TP_{TPM} \sin\theta) < 0.75$ or $> 3.5$, are plotted as open symbols. The dashed lines bound the range in which the thermophysically derived thermal inertia is within a factor of 2 of the estimated value. Error bars have been omitted for the sake of clarity: on the basis of their quoted errors the mean fractional error in the case of the Hanuš et al. (2018) data is 40%; in the case of the MacLennan & Emery (2019) data it is ~ 80%. SI units of thermal inertia are J m$^{-2}$s$^{-0.5}$ K$^{-1}$.

In this work the values of $(TP \sin\theta)$ used to test whether a result from the estimator lies within its range of applicability are based on the thermophysically determined thermal-inertia values where available, which, despite the published large uncertainties, we assume to be relatively reliable compared to values derived using the estimator. A problem for the user of the estimator in general is that an inaccurate thermal-inertia estimate may shift $(TP_{est} \sin\theta)$, where $TP_{est}$ denotes the thermal parameter based on the estimated thermal inertia, into the range of applicability, leading to a false assumption regarding the reliability of the estimate. We discuss this issue in detail below.

## 4. The Effect of Surface Roughness

The thermal-inertia estimator is based on an approximate linear relation between best-fit $\eta$ values and the thermal parameter. A set of thermophysically derived thermal-inertia values for NEOs was used to determine the parameters of the linear relationship (see Harris & Drube 2016, Figure 3), due to the fact that, compared to MBAs, their $\eta$ values cover a much larger range, and observational circumstances, such as aspect angle range, facilitate more reliable thermophysical modeling. Furthermore, infrared observations of large MBAs may suffer from saturation issues (Masiero et al. 2011) and consequently provide less reliable $\eta$ values. As discussed by Harris & Drube (2016), best-fit $\eta$ values depend on surface roughness in addition to thermal inertia. A linear fit to $\eta$/thermal parameter data derived from a set of objects having different surface roughness characteristics to the objects used in the derivation of Equation (1) would have a modified intercept and slope, with rougher/smoother surfaces giving rise to a slightly steeper/shallower slope. Implicit in the use of Equation (1) is the assumption that the surface roughness of the asteroid in question is compatible with the slope of the linear fit.

Hanuš et al. (2018) state that they derive unexpectedly low thermal-inertia values for some asteroids with diameters in the range 10 – 50 km, indicating very fine and mature regolith, and suggest that the existence of a fine-grained regolith layer may be due to a lack of recent bombardment of the surface (impacts would be expected to eject small grains from the surface). If the surfaces of such objects have fewer craters and a dusty regolith layer their relative smoothness could render the thermal-inertia estimator inaccurate. Surface roughness on all spatial scales causes beaming of thermal-infrared emission in the direction of the Sun, which increases the color temperature observed at low phase angles, leading to lower $\eta$ values (see Harris & Lagerros 2002, and references therein). A lack of roughness leads to higher $\eta$ values, which are otherwise indicative of higher thermal inertia. Therefore, the thermal-inertia estimator will produce overestimates in cases in which surface roughness is lower than that of the objects used to define the linear relation between $\eta$ and thermal parameter used in the derivation of Equation (1). The distribution of the Hanuš et al. (2018) and the MacLennan & Emery (2019) data on a plot of $\eta$ versus $TP_{TPM} \sin \theta$ is shown in Figure 4 (cf. Harris & Drube, 2016, Figure 3). It is clear in Figure 4 that many points with very low values of thermal parameter lie above the linear relation of Harris & Drube (2016), causing Equation (1) to produce overestimates of thermal inertia.

To test the possibility that surface roughness, or the lack thereof, causes the estimator to overestimate asteroid thermal inertia, we investigated the dependence of the fractional difference between the thermophysically derived thermal-inertia values and those estimated using Equation (1), i.e. $(\Gamma_{TPM}-\Gamma_{est})/\Gamma_{TPM}$, on surface roughness as derived from the thermophysical modeling of MacLennan & Emery (2019) and Hanuš et al. (2018). The measure of surface roughness used by these authors is the mean surface slope, as defined by Hapke (1984). The results are plotted in Figures 5a and 5b for the data of MacLennan & Emery (2019) and Hanuš et al. (2018), respectively. The uncertainties in the mean surface slopes derived from thermophysical modeling are very large, nevertheless the plots in Figure

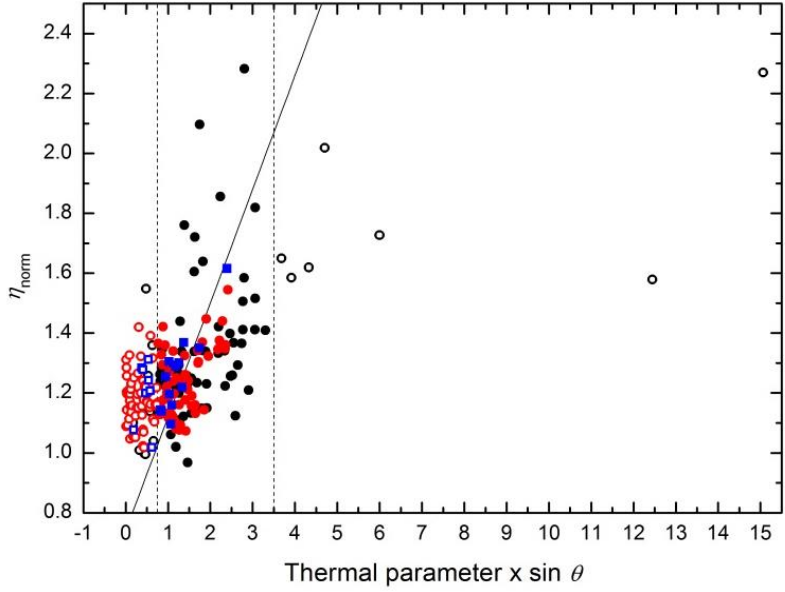

**Figure 4**. Normalized $\eta$ (see Equation (1)) plotted against $TP_{TPM}$ $\sin\theta$ for all three data sets used in this study (Hanuš et al. 2018, Table A.2 – black disks; Hanuš et al. 2018, Table A.3 – red disks; MacLennan & Emery 2019 – blue squares). The vertical dotted lines enclose the range of applicability of the Harris-Drube estimator (0.75 < TP $\sin\theta$ < 3.5); open symbols represent data points with $TP_{TPM}$ $\sin\theta$ outside this range. Objects for which robust $\eta$ values could not be obtained have been excluded. The $\eta$ values have been normalized to a solar phase angle of 50°, as explained in the text. The continuous line is the linear fit forming the basis of the estimator (Equation (1)) given by $\eta_{norm}$ = 0.74 + 0.38 x TP $\sin\theta$. Independent measurements of $\eta$ for the same object are included as separate data points. The two open symbols on the far right of the plot represent (55565) 2002AW$_{197}$, a TNO, and (54509) YORP. The latter has a rotation period of only 12.17 minutes; its very large thermal inertia is probably due to loss of regolith as a result of the high rotation rate. The theoretical $\eta$/TP $\sin\theta$ relation is nonlinear; the linear approximation shown here as a continuous line only applies for TP $\sin\theta$ below about 3.5, i.e. it is not valid for objects with very high thermal-inertia values (for their heliocentric distances) and/or high rotation rates. Note that the uncertainties in the thermal-parameter values (not shown here for the sake of clarity) are dominated by the uncertainties in thermophysically derived thermal inertia and are typically 40% - 80% (see the caption of Figure 3).

5 suggest an association between large fractional differences and low roughness, consistent with the above argument.

It is instructive to note that the largest discrepancies between thermal-inertia values from the Harris-Drube estimator and thermophysical modeling are found for objects in the main asteroid belt with heliocentric distances in the range 2.2 – 3.5 au, as demonstrated in Figure 6. The list of Hanuš et al. (2018, Table A.3) contains very low thermal-inertia values (< 15 J m$^{-2}$s$^{-0.5}$K$^{-1}$) for some 40 out of a total of 122 MBAs. Of these 40 asteroids 38 have heliocentric distances in the range 2.2 – 3.5 au. These authors point out that confirmation of such low thermal-inertia values is very desirable, and in this respect improvements to their thermophysical model together with greater availability of observational data are needed. Nevertheless, for the purposes of this work we take the thermophysically modeled thermal-

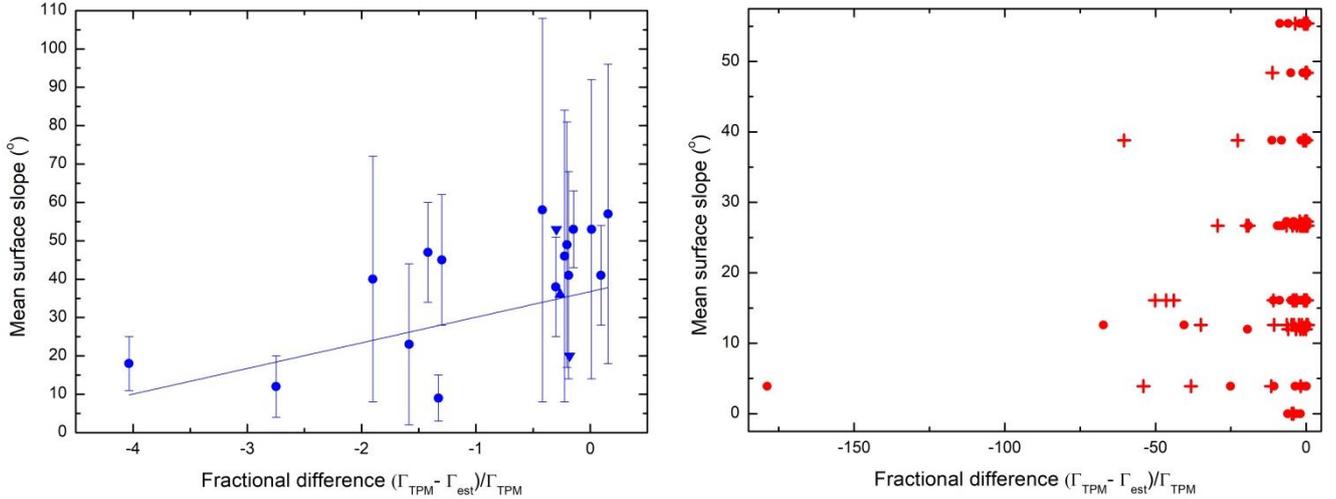

**Figure 5a (left)**. Surface roughness, i.e. mean surface slope as defined by Hapke (1984), plotted against the fractional difference between the thermophysically derived thermal-inertia values of MacLennan & Emery (2019) and those estimated using Equation (1). Triangular symbols indicate upper or lower limits. For objects with two entries in the *WISE* catalog we have taken the mean of thermal-inertia values derived from the Harris-Drube estimator. The weighted linear fit has a positive slope of 6.7 with a standard error of 2.6. The probability of obtaining a similar result or better with uncorrelated variables is less than 3%.

**Figure 5b (right)**. Same as Figure 5a using the thermophysically derived thermal-inertia values of Hanuš et al. (2018). For the objects for which Hanuš et al. (2018) give two thermal-inertia and two mean surface-slope values due to the existence of two pole solutions, we have plotted the second set of results separately using crosses. Small adjustments to the $\Gamma_{est}$ values have been made to normalize them to the heliocentric distances listed by Hanuš et al. (2018) using the $R^{-3/4}$ dependence discussed by Delbo' et al. (2015). Note the fractional differences are much larger than those in Figure 5a.

The uncertainties in mean surface slope are very large as is evident in Figure 5a (Hanuš et al. 2018 do not give uncertainties for their derived mean surface slopes but we assume they are similar). Despite the large uncertainties the plotted data suggest that the largest fractional differences are associated with relatively low surface roughness.

inertia values at face value and alert potential users of the Harris-Drube estimator to the fact that, in its present form, it does not reproduce the very low values derived for some MBAs from the thermophysical modeling of Hanuš et al. (2018). Users of the estimator should be aware that results obtained for MBAs with heliocentric distances in the range 2.2 – 3.5 au may be overestimates.

In the case of NEOs the estimator appears to perform well (Figure 7), with most thermophysically modeled thermal-inertia values lying within a factor of 2 of the estimates. Figure 7 includes the NEOs in Table 1 of Harris & Drube (2016) plus the results published since those listed in Tables 1 and A1.

**Table 2**

Skin Depth (cm) As a Function of Rotation Period and Thermal Inertia

| Thermal Inertia ($J\ m^{-2}s^{-0.5}K^{-1}$) | Rotation Period (hr) | | | | |
|---|---|---|---|---|---|
| | 2 | 10 | 50 | 250 | 1000 |
| 1 | 0.0035 | 0.0079 | 0.018 | 0.039 | 0.079 |
| 5 | 0.018 | 0.039 | 0.088 | 0.20 | 0.39 |
| 15 | 0.053 | 0.12 | 0.26 | 0.59 | 1.2 |
| 100 | 0.35 | 0.79 | 1.8 | 3.9 | 7.9 |
| 1000 | 3.5 | 7.9 | 18 | 39 | 79 |

Note: The calculation of skin depth assumes realistic values of $\rho$ (= 2000 kg $m^{-3}$) and $c$ (= 680 J $kg^{-1}$ $K^{-1}$) (see Harris & Drube 2016, and references therein).

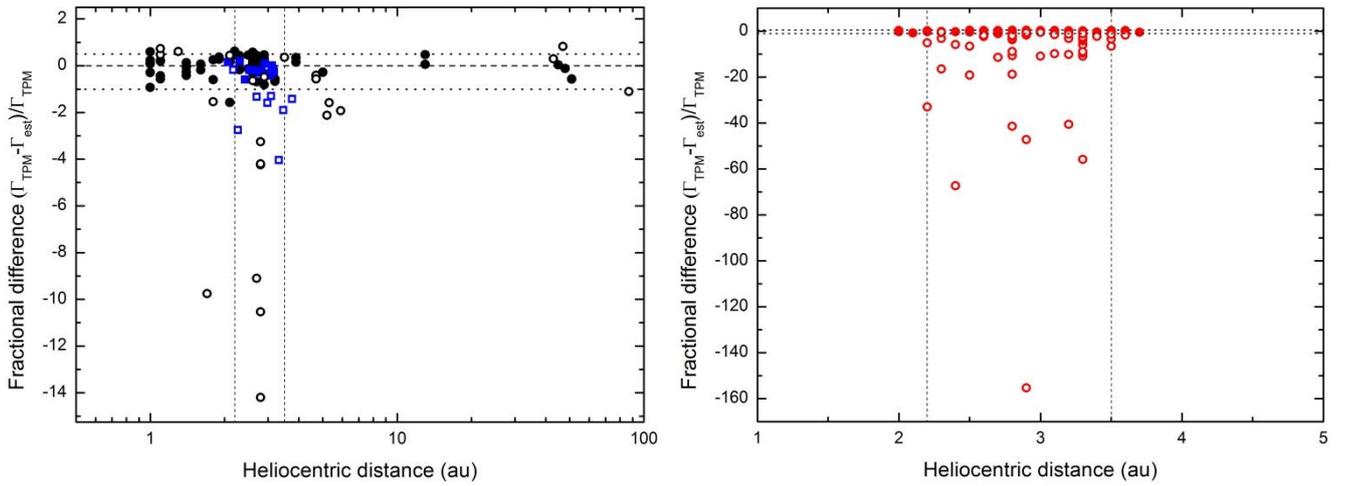

**Figure 6a (left)**. Black disks: the fractional difference between the thermophysically derived thermal-inertia values from the literature (Hanuš et al. 2018, Table A.2) and those estimated using Equation (1) plotted against heliocentric distance. Horizontal dotted lines represent the factor of 2 uncertainty region of the Harris-Drube estimator. Vertical dotted lines enclose the region between 2.2 and 3.5 au. The outlier with $R$ = 1.7 au, fractional difference = -9.75 is (29075) 1950 DA: see Section 6. Values of $\Gamma_{est}$, obtained using the Harris-Drube estimator, have been normalized to the heliocentric distances, $R$, given in Table A.2 of Hanuš et al. (2018) using the $R^{-3/4}$ scaling relation discussed by Delbo' et al. (2015).
Blue squares: as above for the thermophysically derived thermal-inertia values of MacLennan & Emery (2019). For objects with two entries in the *WISE* catalog we have taken the mean of the two corresponding thermal-inertia values derived from the Harris-Drube estimator.
Open symbols denote objects for which the thermophysically derived thermal-inertia values are outside the limits of applicability of the estimator, i.e. those with ($TP_{TPM}$ sin $\theta$) < 0.75 or > 3.5.
**Figure 6b (right)**. Same as Figure 6a using the thermophysically derived thermal-inertia values of Hanuš et al. (2018, Table A.3). For the objects for which Hanuš et al. (2018) give two thermal-inertia values due to the existence of two pole solutions, we have taken the weighted mean for the purposes of this plot. Small adjustments have been made to the estimated thermal-inertia values to normalize them to the heliocentric distances listed by Hanuš et al. (2018, Table A.3).

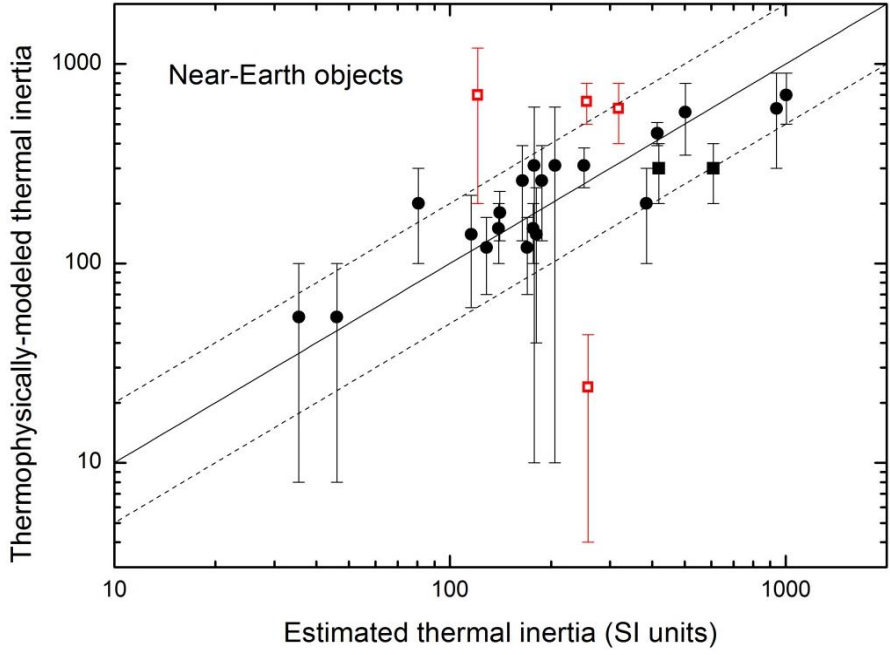

**Figure 7**. Thermophysically modeled thermal inertia for near-Earth objects compared to values derived from the estimator of Harris & Drube (2016). Data are from Harris & Drube (2016) and Table 1, except for objects for which the only thermophysical values available are outside the limits of applicability of the estimator, i.e. those with (TP$_{TPM}$ sin $\theta$) < 0.75 or > 3.5, which are plotted as open red squares and listed in Table A1 (i.e. 3200, 29075, 54509, and 161989). In the case of (162173) Ryugu a "ground-truth" value of thermal inertia derived by Okada et al. (2020) from Hayabusa 2 results has been taken ($\Gamma_{TPM}$ = 300 ± 100 at 1.1 au, plotted as black squares). The outlier near the bottom of the plot is (29075) 1950 DA: see Section 6. Values of $\eta$ for an object derived from independent sets of data are treated as separate values, thus some objects are represented by more than one data point. The dashed lines bound the range in which the thermophysically derived thermal inertia is within a factor of 2 of the estimated value. The estimated, $\Gamma_{est}$, values correspond to the same heliocentric distances, $R$, as the $\Gamma_{TPM}$ values. SI units of thermal inertia are J m$^{-2}$s$^{-0.5}$ K$^{-1}$. In most cases the estimated thermal-inertia values agree within a factor of 2 with those from thermophysical modeling.

## 5. The Effect of Rotation

The depth to which solar energy can penetrate into the surface of an asteroid is a function of thermal inertia and rotation rate. The depth at which the amplitude of the diurnal thermal wave decays to 1/$e$ of its surface value, known as the skin depth, is given by $d_s = (2\kappa/\rho c\omega)^{0.5} = (2/\omega)^{0.5}\Gamma/(\rho c)$ (Wesselink 1948; Spencer et al. 1989), where $\kappa$ is the thermal conductivity, $\rho$ is the density, and $c$ is the specific heat of the surface material. The skin depth increases with rotation period due to the fact that on a slowly rotating asteroid surface elements on the day side have more time to absorb solar energy, which consequently penetrates deeper into the object's surface, depending on the thermal conductivity and therefore the thermal inertia. In Table 2 values of skin depth are listed for various values of rotation period and thermal inertia, assuming realistic values of $\rho$ (= 2000 kg m$^{-3}$) and $c$ (= 680 J kg$^{-1}$ K$^{-1}$) (see Harris & Drube 2016, and references therein). It is clear from Table 2 that for very low values of thermal inertia (< 15 J m$^{-2}$s$^{-0.5}$K$^{-1}$) the skin depth remains below a few millimeters for rotation periods of up to 200 hr. In the case of such low values of thermal inertia, energy from insolation is confined to the topmost layer of regolith even for the highest rotation periods. Only in cases of surface thermal inertia exceeding some 50 J m$^{-2}$s$^{-0.5}$K$^{-1}$ can solar energy penetrate deeper to subsurface material, which may have higher thermal inertia.

The possibility that thermal inertia increases rapidly with depth offers an explanation for the significant increase of thermal inertia with rotation period demonstrated by Harris & Drube (2016). In the light of the discussion in Section 4 we have revisited the question of rotation-dependent thermal inertia in *WISE* data for MBAs. At first sight the results of Marciniak et al. (2019) appear inconsistent with the results of Harris & Drube (2016) due to a number of slowly rotating MBAs in their data set having very low thermal inertia. Taking an extreme case from the target asteroids of Marciniak et al. (2019) as an illustration, the authors derive a thermal inertia value of 20 (+25, -20) J m$^{-2}$s$^{-0.5}$K$^{-1}$ for (538) Friederike, which has a rotation rate of 46.7 hr. Assuming values for density and specific heat as above, the corresponding skin depth is 3.4 mm. Even for the slowest rotator in the dataset of Hanuš et al. (2018), (1424) Sundmania, with a period of 94.5 hr and thermal inertia = 15 (+15, -15) J m$^{-2}$s$^{-0.5}$K$^{-1}$, the skin depth is only 3.6 mm. In the context of the explanation given above, we maintain that in cases of such low skin depths the thermal wave may not penetrate sufficiently, even with relatively long rotation periods, to sample surface material of higher thermal inertia. Biele et al. (2019) have modeled the effects of a thin covering of relatively low thermal-inertia material on the surface temperatures and thermal inertia of an asteroid and find that the thermal inertia of the underlying material can be completely masked by a layer of around 1 skin depth.

In summary, the very low values of thermal inertia found via thermophysical modeling for some MBAs, such as those mentioned above and plotted in Figure 3, are presumably due to a surface covering of fine-grained insulating material (Hanuš et al. 2018). Even if all MBAs had roughly the same thickness of similar insulating material, the skin depth would be different for each object due to its dependence on rotation rate (see above). Therefore, above a particular rotation period the thermal wave would be expected to penetrate down to material with higher thermal inertia, leading to a dependence of observed thermal inertia on rotation period. However, within an overall trend of increasing thermal inertia with rotation period, we would expect exceptions in the cases of asteroids with very low thermal inertia. Future thermal-infrared observations of slowly rotating asteroids, such as those described by Marciniak et al. (2019), together with detailed thermophysical modeling, should provide information on the thickness of the fine-grained insulating layers on MBAs (see Table 2). Figure 8 is a plot of estimated thermal inertia versus rotation period in which objects with $d_s$ < 1 cm and $d_s$ > 1 cm are plotted with different symbols. All objects with (TP$_{est}$ sin $\theta$) outside the range 0.75 - 3.5 and heliocentric distance between 2.2 and 3.5 au were removed from the dataset in an attempt to exclude less accurate estimates. The estimated thermal-inertia values suggest a trend of increasing thermal inertia with increasing period, but it remains an important task to further investigate this phenomenon using detailed thermophysical modeling.

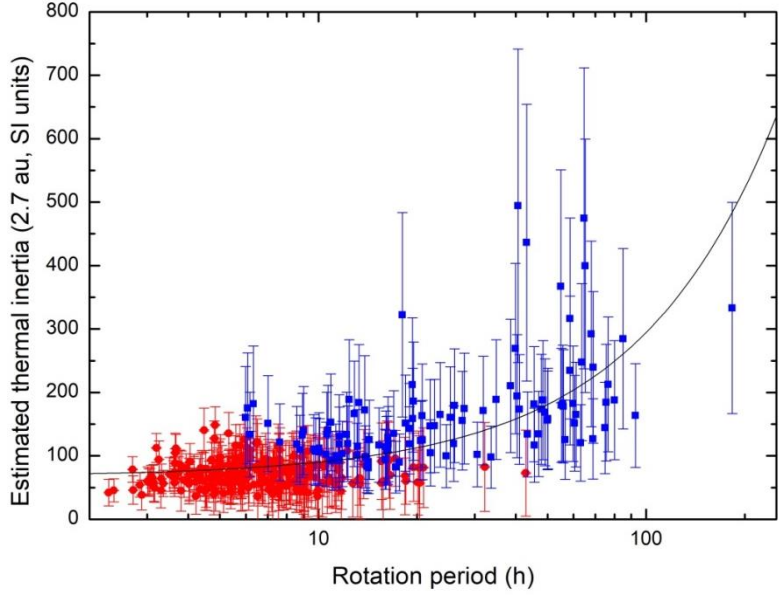

**Figure 8**. Estimated thermal inertia versus rotation period for main-belt asteroids. Thermal-inertia values have been normalized to a heliocentric distance of 2.7 au using the $R^{-3/4}$ scaling relation discussed by Delbo' et al. (2015). Red disks and blue squares denote objects with $d_s < 1$ cm and $d_s > 1$ cm, respectively, where $d_s$ is the skin depth. Objects for which (TP$_{est}$ sin$\theta$) < 0.75 or > 3.5 and those with heliocentric distances in the range 2.2 - 3.5 au have been excluded (see the text). Error bars are ±50%, based on the rms fractional deviation of the data within the range of applicability of the estimator in Figure 1. SI units of thermal inertia are J m$^{-2}$s$^{-0.5}$ K$^{-1}$. The continuous line represents a linear fit (note the logarithmic scale) with a correlation coefficient of 0.67. Relatively high values of thermal inertia are evident amongst objects with long rotation periods and therefore larger skin depths.

## 6. Discussion

### 6.1. Improving the Performance of the Estimator

A further consideration regarding the overestimation of thermal inertia by Equation (1) in the case of smooth surfaces is the normalization of $\eta$ to a reference phase angle (see Section 2). Best-fit $\eta$ values show a general dependence on solar phase angle, as discussed in Section 2. Rough surfaces cause "beaming" of thermal radiation in the solar direction. The resulting increase in color temperature observed at low solar phase angles is accompanied by a reduction in observed color temperature at high phase angles, and thus a corresponding increase in $\eta$. In the case of smoother surfaces this effect would be reduced. We attempted to reduce the fractional differences, ($\Gamma_{TPM}$-$\Gamma_{est}$)/$\Gamma_{TPM}$, as plotted in Figures 5 and 6 by taking the original published $\eta$ values for input into Equation (1) instead of the values normalized to a solar phase angle of 50°. While the fractional differences were reduced in some cases there was no apparent overall improvement in the agreement between the estimated values of thermal inertia and those from thermophysical modeling. Improvements in the performance of the estimator may be possible, at least for certain classes of asteroids, by adjusting the parameters in Equation (1), or attempting to fit more realistic nonlinear functions to the $\eta$ versus thermal parameter data (Figure 4). We have experimented with some functions but have been unable to achieve convincing overall improvements in resulting thermal-inertia estimates. Thermal-inertia values from thermophysical modeling published to date often have uncertainties exceeding 50% (e.g., see Tables 1 and A1), leading to correspondingly uncertain thermal-parameter values. Further attempts at improving the estimator should probably await more reliable results from thermophysical modeling.

## 6.2. The Puzzling Case of (29075) 1950 DA

The case of the potentially hazardous NEO (29075) 1950 DA is instructive. The thermal inertia according to Rozitis et al. (2014) from thermophysical modeling, based on *WISE* observations obtained on 2010 July 12-13, is 24 (+20, -14) J m$^{-2}$s$^{-0.5}$K$^{-1}$. The *WISE* data archive (Mainzer et al. 2019) gives diameter, $D = 2 \pm 0.2$ km and $p_V = 0.07 \pm 0.02$, with $H = 17$, $G = 0.15$, and $\eta = 2.513 \pm 0.386$. The heliocentric distance at the time of the *WISE* observations was 1.74 au. The corresponding thermal-inertia value from the Harris-Drube estimator is 250 J m$^{-2}$s$^{-0.5}$K$^{-1}$, which is an order of magnitude higher than the thermophysically derived value and more in line with values found for other small asteroids (for example, see Delbo' et al. 2015, Figure 9). However, the diameter given in the *WISE* archive (Mainzer et al. 2019) is inconsistent with the values 1.16 and 1.30 km derived by Busch et al. (2007) from radar observations, assuming prograde and retrograde rotation, respectively. How can these discrepant results be reconciled?

Rozitis et al. (2014) argue that observations of Yarkovsky drift imply that the rotation of (29075) 1950 DA must be retrograde and adopt a diameter of $1.3 \pm 0.13$ km. The *WISE* flux measurements used in the modeling of Rozitis et al. (2014) were all taken in the W3 band (12 μm), except for two observations in the W4 band (22 μm). In the Extended Data Figure 1 of Rozitis et al. (2014) their thermophysical model flux predictions are compared with the *WISE* observations; it is evident that the *WISE* W4-band fluxes are 50% higher than predicted by the model. Furthermore, two W3-band fluxes measured simultaneously with the W4-band fluxes are 20% higher than predicted by the thermophysical model of Rozitis et al. (2014). In both cases reducing the *WISE* fluxes by their 1σ uncertainties brings them nearly in line with the model predictions. Taking the simultaneously measured W3- and W4-band fluxes and using the NEATM (Harris 1998), we obtain $D = 1.95$ km, $p_V = 0.074$, and $\eta = 2.24$, very similar results to those quoted above from the *WISE* archive (Mainzer et al. 2019). If we reduce the fluxes to be consistent with the thermophysical model flux predictions of Rozitis et al. (2014, Extended Data Figure 1), the NEATM gives $D = 1.28$ km, $p_V = 0.20$, and $\eta = 1.19$. The diameter and albedo values agree very well with those expected from the thermophysical model of Rozitis et al. (2014). The corresponding thermal-inertia value from the Harris-Drube estimator is 75 (+75, -37) Jm$^{-2}$s$^{-0.5}$K$^{-1}$, which is still higher than the value of 24 (+20, -14) J m$^{-2}$s$^{-0.5}$K$^{-1}$ obtained by Rozitis et al. (2014), but now consistent given the large uncertainties. Furthermore, based on the 10% uncertainty on the radar-derived diameter used in their thermophysical model, Rozitis et al. (2014) give an upper limit of 82 J m$^{-2}$s$^{-0.5}$K$^{-1}$ for the thermal inertia of (29075) 1950 DA.

In summary, taking the *WISE* measurements of (29075) 1950 DA at face value, the Harris-Drube estimator gives a value of thermal inertia that is a factor of 10 higher than that derived by Rozitis et al. (2014) from thermophysical modeling. On the other hand, if we take the *WISE* W3- and W4-band fluxes predicted by the thermophysical model of Rozitis et al. (2014), and derive a best-fit beaming parameter, $\eta$, using the NEATM, the Harris-Drube estimator gives a much lower thermal-inertia value that is compatible with that of Rozitis et al. (2014) given the uncertainties. In the absence of more extensive infrared and radar observations we cannot provide further insight into the thermal inertia of (29075) 1950 DA. However, we have shown that the combination of the NEATM and the Harris-Drube estimator provide consistent physical parameters given the data and modeling results currently available.

### *6.3. Comparison with In Situ Thermal-Inertia Measurements*

Here we provide an example of how the estimator may be used in practice to derive useful information from a large set of fitted $\eta$ values. The surprisingly similar thermal-inertia values found from in situ measurements of rocks on the C-complex NEOs (101955) Bennu and (162173) Ryugu (some 300 - 350 SI units, Lauretta et al. 2019; Okada et al. 2020) are much lower than those expected for solid rock. However, the measured values are higher than would be expected if a significant dust layer were present and are suggestive of highly porous rocky material. Is it reasonable to speculate that the surface characteristics of Bennu and Ryugu may be representative of C-complex asteroids in the main belt? The wide range of thermal-inertia values found amongst MBAs would certainly allow the possibility that some, at least, have surfaces similar to those of Bennu and Ryugu. We filtered the *WISE/NEOWISE* catalog for C-complex (C and B) and S-complex (for comparison) MBAs based on the SMASSII taxonomic system (Bus & Binzel 2002). In order to facilitate comparison of thermal-inertia values from the estimator with the in situ measurements made at heliocentric distances of $R \sim 1$ au, we normalized the estimated thermal-inertia values to $R = 1$ au assuming an $R^{-3/4}$ dependence. The resulting thermal-inertia distributions are shown in Figure 9. While the peak thermal-inertia values are much lower than the in situ measurements, the broad distributions extend out to the values found for Bennu and Ryugu. Note that the median diameters of the objects plotted in Figure 9 are much larger than the km-sized NEOs Bennu and Ryugu (15 km and 57 km for the S- and C-complex objects, respectively). The distribution for the C, B types is slightly broader than that of the S types and drops abruptly at around 300 SI units. The generally lower values estimated for the MBAs compared to the in situ measurements are indicative of layers of fine-grained porous insulating material, e.g. fine dust, as is found on the Moon. A much larger proportion of S-complex objects have thermal-inertia values below 120 SI units, and the C-complex objects appear to be associated with higher thermal inertia in general, suggestive of different thicknesses or properties of fine-grained material on S- and C, B-type asteroids. However, the broad range of thermal inertia extending to the relatively large values seen in Figure 9 suggest that a significant number of C-complex objects have only thin layers of insulating material, or perhaps largely bare surfaces with exposed highly porous, rocky material as found on Bennu and Ryugu.

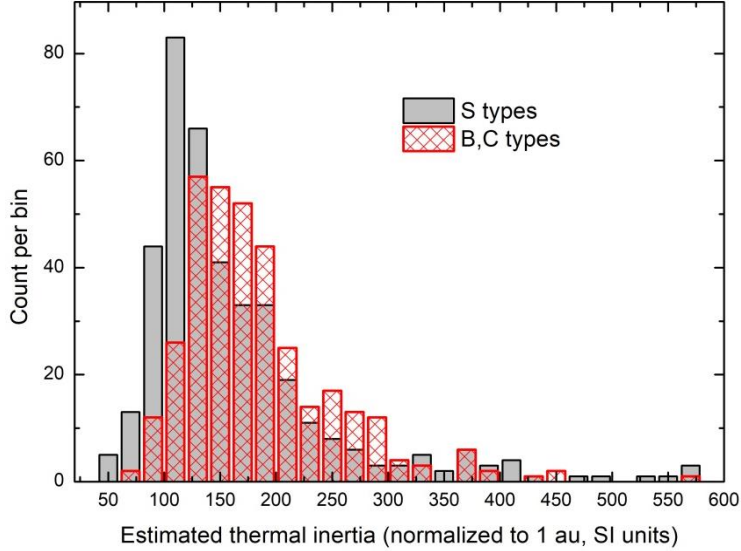

**Figure 9**. Frequency distributions of estimated thermal inertia for main-belt asteroids based on fitted $\eta$ values from the *WISE* catalog (Mainzer et al. 2019). Gray columns denote S-complex asteroids; red hashed columns denote C-complex asteroids. Thermal-inertia values have been normalized to a heliocentric distance of 1 au using the $R^{-3/4}$ scaling relation. Objects for which (TP$_{est}$ sin$\theta$) < 0.75 or > 3.5 have been excluded. For objects with no available pole solution the median value of sin $\theta$ = 0.906 has been assumed. SI units of thermal inertia are J m$^{-2}$s$^{-0.5}$ K$^{-1}$.

We emphasize the preliminary nature of these results. Since in situ measurements for objects of other taxonomic classes are not currently available, it would be premature to draw broad conclusions about the implications of remotely sensed thermal-inertia values for asteroid surface properties on the basis of the results for Bennu and Ryugu alone.

### 7. Conclusions

Within its bounds of applicability, the thermal-inertia estimator introduced by Harris & Drube (2016) provides similar values to those derived from thermophysical modeling over several orders of magnitude. However, outside its formal bounds of applicability, values of thermal inertia derived from thermophysical modeling for some asteroids in the main belt are significantly lower than the values given by the estimator. The user of the estimator should be aware that an overestimated thermal-inertia result may shift the thermal parameter, TP, into the range of applicability (0.75 < TPsin$\theta$ < 3.5), leading to a false assumption regarding the reliability of the estimate. A possible explanation for the discrepant results is reduced surface roughness, which increases the $\eta$ values used as input to the estimator thus mimicking the effect of higher thermal inertia. Pending confirmation of the very low thermal-inertia values ($< 15$ J m$^{-2}$s$^{-0.5}$K$^{-1}$) derived via thermophysical modeling for some MBAs (Hanuš et al. 2018), we alert potential users of the estimator to the fact that, in its present form, it does not reproduce the very low values derived for these MBAs, and results obtained for MBAs with heliocentric distances in the range 2.2 – 3.5 au may be overestimates.

Harris & Drube (2016) provided evidence for a dependence of remotely sensed thermal inertia on rotation period, which they interpreted in terms of deeper penetration of solar energy into subsolar surface elements of longer period objects. At first sight the results of Marciniak et al. (2019) appear inconsistent with the results of Harris & Drube (2016) due to a number of slowly rotating MBAs in their dataset having very low measured thermal inertia and therefore very small skin depths. We have shown that in cases of such low thermal inertia the thermal wave may not penetrate sufficiently, even with relatively long rotation periods, to sample subsurface material of higher thermal inertia. Therefore, within an overall trend of increasing thermal inertia with rotation period, exceptions in the cases of asteroids with very low thermal inertia should be expected.

For near-Earth objects the Harris-Drube thermal-inertia estimator gives values that are generally in good agreement with those derived from thermophysical modeling. An exception is (29075) 1950 DA, for which the estimator gives a thermal inertia of 250 Jm$^{-2}$s$^{-0.5}$K$^{-1}$ compared to 24 Jm$^{-2}$s$^{-0.5}$K$^{-1}$ from the thermophysical modeling of Rozitis et al. (2014). An important issue in the case of (29075) 1950 DA is the discrepancy between some *WISE* W3- and W4-band flux measurements and the predictions of the thermophysical model. We find that NEATM fits to two pairs of simultaneous W3 and W4 flux measurements provide results for diameter, albedo, and $\eta$ that are in agreement with those published in the *WISE* archive (Mainzer et al. 2019). Furthermore, the $\eta$ value implies a relatively high thermal-inertia estimate in line with values found for other small asteroids (Delbo' et al. 2015). On the other hand, if the lower flux predictions of the thermophysical model are taken, we obtain results, including a thermal-inertia estimate, in agreement with those of Rozitis et al. (2014). While the resolution of these ambiguous results will require further, higher quality observational data, this example offers a demonstration of how the NEATM/Harris-Drube estimator combination provides reasonable physical parameters given a consistent set of data.

A comparison of estimated-thermal-inertia frequency distributions of C-complex (C and B types) and S-complex MBAs in the *WISE/NEOWISE* catalog reveals that C-complex objects appear to be associated with higher thermal-inertia values in general and have a broader distribution, extending up to values similar to those found from in situ measurements of rocks on the NEOs Bennu and Ryugu. The generally lower values estimated for the MBAs compared to the in situ measurements are indicative of surface layers of fine-grained porous insulating material. Different thicknesses or properties of insulating material in the two cases may explain the generally lower thermal-inertia values found for the S-type MBAs.

The *WISE* mission has provided a vast amount of thermal-infrared data on asteroids and future missions, such as NASA's NEO Surveillance Mission currently under development, promise to provide much more. Given the data requirements and complexity of thermophysical modeling, tools allowing "quick-look" assessment of thermal-infrared data of NEOs can play an important role in preliminary analyses of large sets of thermal-infrared data, checking the consistency of observational results, and guiding the planning of future observations.

We thank the two anonymous reviewers for their insightful comments, which led to significant improvements in the presentation of the work. This publication makes use of data products from *WISE/NEOWISE*, which is a project of the Jet Propulsion Laboratory/California Institute of Technology, funded by the Planetary Science Division of the National Aeronautics and Space Administration. We gratefully acknowledge the excellent JPL Solar System Dynamics web service of which we have made extensive use. This work was initiated with support from the European Union's Horizon 2020 research and innovation program under grant agreement No. 640351 (project NEOShield-2).

**Appendix**

The plots in Figures 1, 2, and 7 include objects for which the thermal-parameter values are outside the formal limits of applicability of the estimator, i.e., those with ($TP_{TPM}$ $\sin\theta$) <0.75 or >3.5. The relevant parameters of these objects are listed in Table A1. Note that for the purposes of Figure 7 only the results of Harris and Davies (1999) are taken for (1980) Tezcatlipoca (see Table 1), due to the corresponding $TP_{TPM}$ $\sin\theta$ values being within the limits of applicability of the estimator.

# Table A1

Data in Figures 1, 2, and 7 for Which (TP$_{TPM}$ sin$\theta$) Is Outside the Limits of Applicability of the Estimator (0.75-3.5)

| Name | G | R (au) | Period (hr) | $p_V$ | α° | sin θ | η | $η_{norm}$ | TP$_{TPM}$ sinθ | $Γ_{est}$ | $Γ_{TPM}$ | Data Source |
|---|---|---|---|---|---|---|---|---|---|---|---|---|
| 2 Pallas | 0.11 | 3.25 | 7.813 | 0.14 | <20 | 0.82 | 0.85±0.06 | 1.13±0.07 | 0.22 | 46.5 | 10±10 | 1 |
| 3 Juno | 0.32 | 2.20 | 7.21 | 0.22 | 27.3 | 0.97 | 0.88±0.10 | 1.10±0.13 | 0.08 | 58.9 | 5±5 | 1 |
| 21 Lutetia | 0.11 | 2.83 | 8.165 | 0.20 | 20.8 | 0.99 | 0.93±0.07 | 1.22±0.10 | 0.11 | 57.2 | 5±5 | 1 |
| 21 Lutetia | 0.11 | 2.06 | 8.165 | 0.19 | 27.4 | 0.82 | 0.93±0.19 | 1.15±0.23 | 0.06 | 95.5 | 5±5 | 2 |
| 45 Eugenia | 0.07 | 2.53 | 5.699 | 0.04 | 23.3 | 0.66 | 0.88±0.08 | 1.14±0.10 | 0.62 | 74.5 | 45±45 | 1 |
| 45 Eugenia | 0.07 | 2.55 | 5.699 | 0.04 | 23.2 | 0.60 | 0.88±0.01 | 1.14±0.01 | 0.57 | 81.5 | 45±45 | 3 |
| 121 Hermione | 0.15 | 3.00 | 5.551 | 0.05 | <20 | 0.55 | 0.71±0.10 | 1.00±0.15 | 0.46 | 43.4 | 30±25 | 1 |
| 532 Herculina | 0.15 | 2.70 | 9.405 | 0.17 | 21.5 | 1.00 | 0.78±0.03 | 1.06±0.04 | 0.19 | 43.7 | 10±10 | 1 |
| 532 Herculina | 0.15 | 2.29 | 9.405 | 0.20 | 26.1 | 1.00 | 0.87±0.02 | 1.10±0.03 | 0.15 | 63.2 | 10±10 | 3 |
| 617 Patroclus | 0.15 | 5.95 | 103 | 0.04 | <20 | 1.00 | 0.85±0.17 | 1.14±0.23 | 0.36 | 58.2 | 20±15 | 4 |
| 956 Elisa | 0.15 | 1.83 | 16.49 | 0.14 | 29.3 | 0.91 | 1.16±0.05 | 1.36±0.06 | 0.64 | 226 | 90±60 | 5 |
| 1980 Tezcatlipoca | 0.15 | 2.06 | 7.246 | 0.13 | 29.4 | 0.99 | 1.42±0.04 | 1.62±0.05 | 4.33 | 151 | 310±300 | 3 |
| 1980 Tezcatlipoca | 0.15 | 2.33 | 7.246 | 0.14 | 25.0 | 0.99 | 1.29±0.10 | 1.53±0.12 | 5.17 | 125 | 310±300 | 3 |
| 2606 Odessa | 0.15 | 3.48 | 8.243 | 0.17 | <20 | 1.00 | 1.36±0.02 | 1.65±0.03 | 3.67 | 80.9 | 125±75 | 3 |
| 2867 Steins | 0.15 | 2.43 | 6.049 | 0.30 | 23.8 | 1.00 | 1.27±0.07 | 1.52±0.08 | 4.39 | 97.6 | 210±30 | 3 |
| 3063 Makhaon | 0.15 | 5.45 | 17.3 | 0.05 | <20 | 0.906 | 0.71±0.14 | 0.99±0.2 | 0.52 | 19.0 | 15±15 | 1 |
| 3200 Phaethon | 0.15 | 1.13 | 3.604 | 0.11 | 48.0 | 0.91 | 1.6±0.32 | 1.62±0.32 | 4.42 | 312 | 600±200 | 6 |
| 29075 1950DA | 0.15 | 1.74 | 2.122 | 0.07 | 35.7 | 1.00 | 2.51±0.39 | 2.65±0.41 | 0.49 | 253 | 24±20 | 3 |
| 50000 Quaoar | 0.15 | 43.2 | 17.68 | 0.13 | <20 | 0.906 | 1.73±0.08 | 2.02±0.09 | 4.72 | 4.25 | 6±4 | 8 |
| 54509 YORP | 0.15 | 1.09 | 0.203 | 0.20 | 59.3 | 1.00 | 2.36±0.5 | 2.27±0.48 | 14.96 | 121 | 700±500 | 7 |
| 55565 2002AW$_{197}$ | 0.15 | 46.5 | 8.87 | 0.11 | <20 | 0.906 | 1.29±0.13 | 1.58±0.16 | 12.35 | 1.77 | 10±10 | 8 |
| 90377 Sedna | 0.15 | 87.4 | 10.27 | 0.41 | <20 | 0.906 | 0.72±0.7 | 1.01±0.98 | 0.33 | 0.21 | 0.1±0.1 | 8 |
| 161989 Cacus | 0.15 | 1.32 | 3.754 | 0.20 | 48.1 | 0.906 | 1.71±0.22 | 1.73±0.22 | 6.07 | 276 | 650±150 | 3 |

**Notes:** The listed thermophysically modeled thermal-inertia values, $\Gamma_{TPM}$, are from Hanuš et al (2018, Table A.2). For comparison purposes the listed values of $\Gamma_{est}$, obtained using the Harris & Drube (2016) estimator, have been normalized to the heliocentric distances, $R$, given in Table A.2 of Hanuš et al. (2018) using the $R^{-3/4}$ scaling relation discussed by Delbo' et al. (2015). Units of thermal inertia are $Jm^{-2}s^{-0.5}K^{-1}$. Other symbols are $G$: slope parameter, $p_V$: geometric visual albedo, $\alpha$: solar phase angle, $\eta$: beaming parameter, and $\theta$: angle between the object's spin axis and the solar direction. For objects with no available pole solution, the median value of sin $\theta = 0.906$ has been assumed. Data sources for $G$, $p_V$, $\eta$ are:

1. IRAS SIMPS Catalog (Tedesco et al. 2002), $\eta$ values are from this work;
2. Mueller et al. (2006);
3. *WISE* Catalog (Mainzer et al. 2019);
4. Mueller et al. (2010) ($\Gamma_{TPM}$ is based on eclipse measurements and is expected to be lower than the diurnal value);
5. Lim et al. (2011);
6. Harris (1998);
7. Mueller (2007);
8. Lellouch et al. (2013).

Observing geometry was taken from the NASA Jet Propulsion Laboratory Solar System Dynamics Horizons facility and asteroid spin vectors from the Asteroid Lightcurve Database (Warner et al. 2018).

## ORCID iDs

Alan W. Harris https://orcid.org/0000-0001-8548-8268
Line Drube https://orcid.org/0000-0003-2486-8894